\documentclass[a4paper,11pt]{article}
\usepackage{jcappub} 
\arxivnumber{2502.20962} 

\newcommand{\hi}{H{\sc i}}

\title{\boldmath Impact of Calibration and Position Errors on Astrophysical Parameters of the  {\hi} 21cm Signal}

\author[a,1]{Anshuman Tripathi,\note{Corresponding author.}}
\author[a]{Abhirup Datta,}
\author[b]{Aishrila Mazumder,}
\author[a,c]{Suman Majumdar}

\affiliation[a]{Department of Astronomy, Astrophysics and Space Engineering, Indian Institute of Tech-
nology Indore, Indore, India – 453552.}
\affiliation[b]{Jodrell Bank Centre for Astrophysics, Department of Physics and Astronomy, The University of Manchester, Manchester M13 9PL, UK}
\affiliation[c]{ Department of Physics, Blackett Laboratory, Imperial College, London SW7 2AZ, U. K}

\emailAdd{anshumantripathi85@gmail.com}

\abstract{The Epoch of Reionization (EoR) and Cosmic Dawn (CD) are pivotal stages during the first billion years of the universe, exerting a significant influence on the development of cosmic structure. The detection of the redshifted 21-cm signal from these epochs is challenging due to the dominance of significantly stronger astrophysical foregrounds and the presence of systematics. This work used the 21cm E2E (end to end) pipeline, followed by simulation methodology described \cite{2022Mazumder} to conduct synthetic observations of a simulated sky model that includes both the redshifted 21-cm signal and foregrounds. A framework was constructed using Artificial Neural Networks (ANN) and Bayesian techniques to directly deduce astrophysical parameters from the measured power spectrum. This approach eliminates the need for explicit telescope layout effects correction in interferometric arrays such as SKA-Low. The present work investigates the impact of gain calibration errors and sky model position errors on the recovery of the redshifted 21-cm power spectrum for the SKA-Low AA$^{\ast}$ array configuration. We assessed the effects of these inaccuracies on the deduced astrophysical parameters and established acceptable tolerance levels. Based on our results, the gain calibration error tolerance for ideal signal detection is 0.001 \%. However, if the sky model position errors exceed 0.048 arcseconds, the remaining foregrounds would obscure the target signal.}

\begin{document}
\maketitle
\flushbottom

\section{Introduction}
\label{sec:intro}

The redshifted 21cm line is a potential probe of the early Universe \citep{Field1958, Field1959a, Field1959b}, mainly from the era post recombination until the universe became fully ionized. Based on the theoretical model, in the first billion years of the Universe, Cosmic Dawn (CD) is when the first star or galaxy will be formed (30 $> z > $ 12). These stars and galaxies are formed due to gravitational instability, which causes small-scale fluctuation in the matter density. The UV photons produced by these objects ionized the neutral hydrogen ({\hi}) in the intergalactic medium (IGM). This transition period is known as Epoch of Reionization (EoR) \citep{Furlanetto2006, morales2010, Pritchard2012, BARKANA20161, DAYAL20181}. Based on various indirect observations such as quasar absorption \citep{fan2006} at high redshifts, Thompson scattering optical depth \citep{Planck2018I} suggest that the reionization extended the process and lasted at least til the redshift $z \sim 6$. 

The observation of {\hi} 21cm power spectrum using large interferometric arrays currently holds the most significant potential to observe the redshifted {\hi} 21cm line \citep{Bharadwaj2001, Bharadwaj2005, Morales2005, Zaldarriaga2004}. The 21cm signal inherently encodes information about the underlying dark matter distribution and the properties of the ionizing sources. As a result, it has the potential to trace the history of reionization, reflecting the evolution of the average ionization state of the intergalactic medium (IGM) with redshift during the EoR. Detection of the redshifted {\hi} signal is the key science goal of several ongoing and future experiments. Recently, the detection of a global 21cm signal reported by the Experiment to Detect Global Epoch of Reionization Signature (EDGES) team \citep{Bowman2018}. However, the detection has been challenged by another independent experiment, SARAS \citep{saras,Saurabh2022}. Besides EDGES and SARAS, there are other independent single radiometer experiments like BIGHORNS \citep{bighorns}, SCI-HI \citep{sci-hi}, and LEDA \citep{leda} are also aiming to detect the global signal but have not yet to report any detection. Conversely, interferometers focusing on statistical fluctuations have yielded significant upper limits on the EoR power spectrum (PS) amplitude. The most sensitive operational interferometers, such as the GMRT, MWA, LOFAR, and HERA, have all established upper limits on the power spectrum amplitude of the signal \citep{pacgia,mwa2, 2024Acharya, 2025Mertens_upper_limit, 2023Hera_upper_limit}, but there has been no confirmed detection of the cosmological \hi\ 21cm signal.

However, observing the 21cm signal is highly challenging due to bright astrophysical foregrounds primarily Galactic synchrotron emission and extragalactic point sources that are several orders of magnitude brighter than the signal of interest \citep{jelic2008, jelic2010, Chapman2015, 2004_Di_Matteo,samir, arnab2, aishrila2020}. Additionally, other sources of contamination, such as the Earth's ionosphere and instrument systematics, make detection even more difficult. The observation of the 21cm signal heavily relies on the accuracy of foreground removal and the use of instruments with high sensitivity and controlled systematics. Over the past decade, several novel methods have been proposed to quantify and mitigate each type of contamination for foreground avoidance and removal \citep{Datta2010,Trott2012,chapman2014,Vedantham2012,Thyagarajan2015a,Thyagarajan2015b,mertens2018,hothi} , model the systematics of the instruments \citep{delera2017,trott2017,Joseph2018,Li2018} , address calibration effects \citep{offringa2015,barry16,trottwayth2016,ewall2017,Patil2017,dillon18,Kern2019,jais}, and account for the impact of the Earth's ionosphere \citep{jordan2017,Trott2018, 2024Pal}. These advancements are paving the way for highly sensitive next-generation interferometers, such as the Hydrogen Epoch of Reionization Array (HERA, \citep{DeBoer2017}) and the Square Kilometer Array (SKA-Low, \citep{2015Koopmans}), to detect the 21-cm signal and characterize the multi-redshift power spectrum (PS). This will result in tighter constraints on astrophysical parameters in the early universe. The forthcoming SKA-Low is specifically designed to have the sensitivity needed to detect the PS precisely and is anticipated to generate tomographic images of the HII regions \citep{Mellema2014}.

The EoR signal can be distinguished from foreground contamination because the EoR signal exhibits a spectral structure and is inherently uniform in spatial wavenumber (k) space, whereas foregrounds are spectrally smooth \citep{Morales2004}. The foregrounds' smooth spectral nature, along with the inherent chromaticity of the instruments, limit the contamination to the 'wedge' in cylindrical Fourier space (i.e., the 2D power spectrum) \citep{Datta2010, Trott2012, Vedantham2012}.  The area outside this wedge, where the foregrounds are less prominent than the EoR signal, is referred to as the 'EoR window' \citep{Morales2012}. However, the interaction between astrophysical foregrounds and the instrument results in the leakage of wedge power into the clean modes of the window, a phenomenon known as 'mode mixing' \citep{Morales2012}. This mode mixing impacts the "EoR window," the region outside the wedge, complicating the detection process. One major recurring challenge in detecting cosmic signals is improper calibration. In radio astronomy sky-based calibration is commonly used, but CD/EoR observations often produce inaccurate models due to low angular resolutions and noise confusion, resulting in residual errors and hindering target cosmological signal detection \citep{Datta2009, Datta2010, 2016Barry, ewall2017}. The redundant calibration method, investigated by observatories such as HERA, addresses this by repeatedly measuring the redundant baselines of interferometers to correct for the incoming sky signal and instrumental parameters \citep{2016Dillon}. However, as demonstrated by \citep{Byrne2019}, redundant calibration remains susceptible to errors introduced during the absolute calibration step, which necessarily depends on a sky model. Even in the case of a perfectly redundant array with identical antenna beams, incompleteness in the sky model leads to frequency-dependent calibration errors that can contaminate the 21cm power spectrum. To mitigate these limitations, \citep{2021Byrne} introduced a unified Bayesian calibration framework that integrates both sky-based and redundant approaches. This framework explicitly accounts for instrumental systematics, such as antenna position offsets, beam non-uniformities, and incomplete sky models, thereby improving the robustness and accuracy of 21cm power spectrum estimation.


The signal-to-noise ratio (SNR) is a critical factor in the detection of faint cosmic signals, such as the redshifted 21-cm line from the epoch of reionization. \textbf{\cite{Datta2010}} demonstrated that even minor calibration errors can significantly reduce the dynamic range, thereby obscuring the cosmological signal. Building on this, \cite{2022Mazumder}, through an analysis of the one-dimensional power spectrum (1D PS), determined that the optimal calibration error tolerance for reliable signal detection is approximately $0.01\%$. Consistently, \cite{2016Barry} showed that calibration errors must be limited to below approximately $10^{-5}$ or 0.001 \% in amplitude in order to prevent contamination of the EoR window in power spectrum measurements. Despite these insights, further investigation is needed to determine the precise tolerance limits for various instrumental imperfections that could hinder weak signal detection. This study aims to quantify the tolerance levels required for the successful detection of the redshifted 21-cm signal and the recovery of astrophysical parameters from the epoch of reionization using the SKA-Low telescope. To achieve this, we utilize a hybrid machine learning (ML) approach that integrates artificial neural networks (ANN) with Bayesian methods to analyze the effects of imperfections on observational data and their direct impact on the associated astrophysical parameters. 

Over the past few years, machine learning (ML) techniques have seen extensive application in various areas of cosmology and astrophysics, particularly in imaging, statistics and inference. Among these ML techniques, artificial neural networks (ANNs) are commonly used in 21cm cosmology for signal modelling, in both kinds,  the global signal \citep{cohen2020emulating, globalemu, VAE} and the power spectrum \citep{2018Pritchard, 2022Himanshu}. ANNs are also employed to infer parameters linked to the signal directly, bypassing traditional Bayesian approaches in both the global 21cm signal \citep{choudhury2020extracting, choudhury2021using, Tripathi2024, 2024Tripathi_samp} and power spectrum \citep{shimabukuro2017analysing, Choudhury_PS_2022}. Besides ANNs, other ML algorithms are also widely used in various applications of 21cm cosmology. For example, \citep{hassan2019identifying} utilized Convolutional Neural Networks (CNNs) to detect reionization sources in 21-cm maps. In \citep{chardin2019deep}, deep learning models were employed to replicate the entire time-evolving 21-cm brightness temperature maps from the reionization epoch. The authors validated their predicted 21-cm maps against brightness temperature maps generated by radiative transfer simulations. \cite{gillet2019deep} employed deep learning with CNNs to directly extract astrophysical parameters from 21-cm images. \cite{jennings2019evaluating} conducted a comparative analysis of machine learning techniques for predicting the 21-cm power spectrum from reionization simulations. \cite{Li2019} proposes a Convolutional Denoising Autoencoder (CDAE) to recover the Epoch of Reionization (EoR) signal by training on SKA images simulated with realistic beam effects.

In this work, we have developed an emulator using an artificial neural network (ANN) framework. This trained ANN emulator was employed as model statistics to constrain EoR astrophysical parameters from the total observed sky power spectrum, which includes the {\hi} signal and systematic effects, via a Bayesian inference process. The motivation for building ANN emulators arises from the computational challenges in EoR 21cm cosmology. Generating numerous observable model signals for multi-dimensional parameter space to perform Bayesian inference using semi-numerical or radiative transfer methods is computationally intensive. Additionally, incorporating telescope layout effects through simulated observations further increases the computational expense. To address these challenges, we adopted a formalism already utilized by several groups \citep{2018Pritchard, 2022Himanshu}, using emulators for EoR signal modeling instead of actual simulations. To construct the training datasets for the emulators, we performed simulated radio interferometric observations using a 21cm E2E pipeline for SKA-Low core  array configurations. This allowed us to calculate the total observed sky power spectrum with telescope layout effect. We also studied the systematic biases introduced in the observed power spectrum by comparing it with the true power spectrum. Furthermore, we examined how these biases influence parameter extraction and quantified the tolerance levels for calibration and position errors necessary for the successful detection of the {\hi} 21cm signal from the EoR using this sensitive telescope through Bayesian inferences. This is followup work by \cite{2022Mazumder}, which quantified the tolerance for calibration and position errors in detecting the {\hi} 21cm signal using various interferometers by analyzing variations in RMS in the image plane and visually examining the 2D and 1D power spectra. Our research investigates the effects of gain calibration and position errors on astrophysical parameters, quantifying the tolerance levels needed to obtain inferred parameters that closely match the true values. Specifically, this work highlights the impact of these errors on the 21cm signal’s astrophysical parameters when no mitigation techniques are applied to the residual foreground contamination caused by them. Furthermore, we are developing a mitigation pipeline, as outlined by \textit{Beohar E. et al., in prep}, to effectively correct these errors, enabling the accurate inference of the true power spectrum and astrophysical parameters. 

The structure of this paper is organized as follows: Section \ref{sec: sig_fore} outlines the simulation methodology for the {\hi} signal and provides a description of the foreground models used. Section \ref{sec:synthetic obs} provides the input parameters and telescope array information for performing synthetic observations. Section \ref{sec:ps} discusses the Power Spectrum (PS) estimation. Section \ref{sec:emu ps} covers the emulation details of the PS. Section \ref{sec:Bayesian_Inference} presents the formulation of Bayesian Inference for EoR parameter estimation. Section \ref{sec:err_cov} outlines the method for calculating error covariances of the PS. Finally, Section \ref{sec:result} presents the results.

\section{Astrophysical Components in the Simulation}
\label{sec: sig_fore}
To perform the synthetic observation using the 21cm E2E pipeline, the sky model provided to the pipeline includes the simulated redshifted 21cm signal along with the point source astrophysical foreground model, as detailed below.
\subsection{{\hi} 21cm Maps}
\label{sec: HI_sim}
To generate 21cm maps for our simulated observation, we use a semi-numerical simulation 21cmFAST \cite{Mesinger2011, 2020Murray}. 21cmFAST generates {\hi} 21cm maps by first constructing a matter density field and applying the Zeldovich approximation. It employs the guided excursion set formalism to convert the matter density field at a given redshift into an ionization field, which is subsequently used to derive 21-cm brightness temperature fluctuations. In contrast to a detailed radiative transfer approach, this method employs perturbation theory, excursion set formalism, and analytical prescriptions to generate evolved fields for density, ionization, peculiar velocity, and spin temperature. These fields are subsequently integrated to determine the 21-cm brightness temperature. Instead of relying on a halo finder, the code directly processes the evolved density field, enhancing computational efficiency and minimizing memory consumption. This enables the production of numerous realizations of 21-cm maps, power spectra of brightness temperature, matter density, velocity, spin temperature, and ionization fraction at specific redshifts, all at a very low computational cost.

We simulated lightcones in a $500$ $Mpc^{3}$ comoving box with $232^{3}$   of  grid cells for a range for redshift extent of $8.73$ $\leq$ $z$ $\leq$ $9.29$. The constructed lightcone volume is saved on a grid of size 232 × 232 × 64 and projected on a World Coordinate System (WCS) to generate an input signal for simulating observations. 
For this study, we simulate different lightcones to probe different realizations of reionization history using varying EoR parameters. To simulate these lightcones, we use three key EoR parameters $R_{\mathrm{mfp}}$, $T_{\mathrm{vir}}$, and $\zeta$ which can be tweaked to create different reionization histories. These parameters are commonly used to characterize the EoR, as they effectively capture the timing, source population, and morphology of reionization while remaining physically interpretable and computationally efficient \citep{2015Greig}.
We describe these in detail below following \citep{2015Greig, shimabukuro2017analysing}: 
\begin{itemize}
    \item $R_{\mathrm{mfp}}$, Mean Free path of ionizing photon: The ionizing photons travel through the ionized IGM strongly depends on the presence of absorption systems and the sizes of ionized regions \cite{McQuinn2011}. Distance travel by a photon from its source of origin to its sink within ionized regions is called the mean free path of ionizing photon \citep[]{Sobacchi2014}. $R_{\mathrm{mfp}}$ determines the size of the ionized regions.
    \item $\rm T_{\mathrm{vir}}$, Minimum virial temperature of haloes producing ionizing photons: This represents the minimum mass of haloes producing ionizing photons during the EoR. Usually, $\rm T_{\mathrm{vir}}$ is chosen to be  larger than $10^{4}$K such that atomic cooling become effective \citep{Sobacchi2013, 2013Fialkov}.

    \item $\rm \zeta$, Ionizing efficiency: Ionizing efficiency refers to the ability of sources, such as stars or galaxies, to convert their energy into ionizing photons that can ionize hydrogen in the intergalactic medium. This is a combination of several degenerate astrophysical parameters and is defined as $\zeta$ = $f_{\mathrm{esc}}$$f_{\mathrm{\star}}$ $N_{\mathrm{\gamma}}$/(1 + $n_{\mathrm{rec}})$ \citep{2004Furlanetto, 2006Furlanetto}. Here, $ \rm f_{esc}$ is the fraction of ionizing photons escaping from galaxies into the IGM,  $\rm f_{\star}$ is the fraction of baryons locked into stars and $\rm N_{\mathrm{\gamma}}$ is the number of ionizing photons produced per baryon in stars and $\rm n_{\mathrm{rec}}$ is the mean recombination rate per baryon.
\end{itemize}

\subsection{Foreground Models}
\label{sec:Fore}
The foreground component employed in this study includes compact sources based on the Tiered Radio Extragalactic Continuum Simulation (T-RECS, \citep{trecs}). T-RECS simulates the continuum radio sky from 150 MHz to 20 GHz, modeling Active Galactic Nuclei (AGNs) and Star-Forming Galaxies (SFGs), incorporating observational constraints for realistic cosmological evolution. For this study, a subset covering approximately 4 deg$^2$ was extracted from the full 25 deg$^2$ T-RECS catalogue, with a flux cut-off applied between 0.6 Jy to 3.1 mJy at 150 MHz. Flux densities originally specified at 150 MHz were extrapolated to 142 MHz using a spectral index $\alpha = -0.8$, yielding a total of 2522 compact sources within the chosen field of view. To maintain simplicity, the current analysis focuses solely on compact sources and excludes both diffuse foreground emission and complex sources with extended morphology or non-power-law spectral characteristics. Future studies will incorporate these additional components to evaluate their influence on signal recovery.

\section{Synthetic Observations}
\label{sec:synthetic obs}
To conduct synthetic observations, we utilized a 21cm end-to-end (E2E) pipeline \citep{2022Mazumder}. This pipeline employs the OSKAR software package \citep{2009oskar} for simulating SKA-Low configurations, the Common Astronomy Software Applications (CASA) package \citep{2009casa} for further reading and processing of the visibility data. In this simulated observation, the sky was observed with a phase center at $\alpha = 15^\text{h} 00^\text{m} 00^\text{s}$ and $\delta = -30^\circ$ for a duration of 4 hours ($\pm 2$ hours hour angle). The observing bandwidth of the lightcone spans 8 MHz with a channel separation of 125 kHz. For additional details, refer to Tab.~\ref{tab:sim_parms}. 

\subsection{Telescope Model}
\begin{figure*}
    \centering
    {\includegraphics[width=0.5\textwidth]{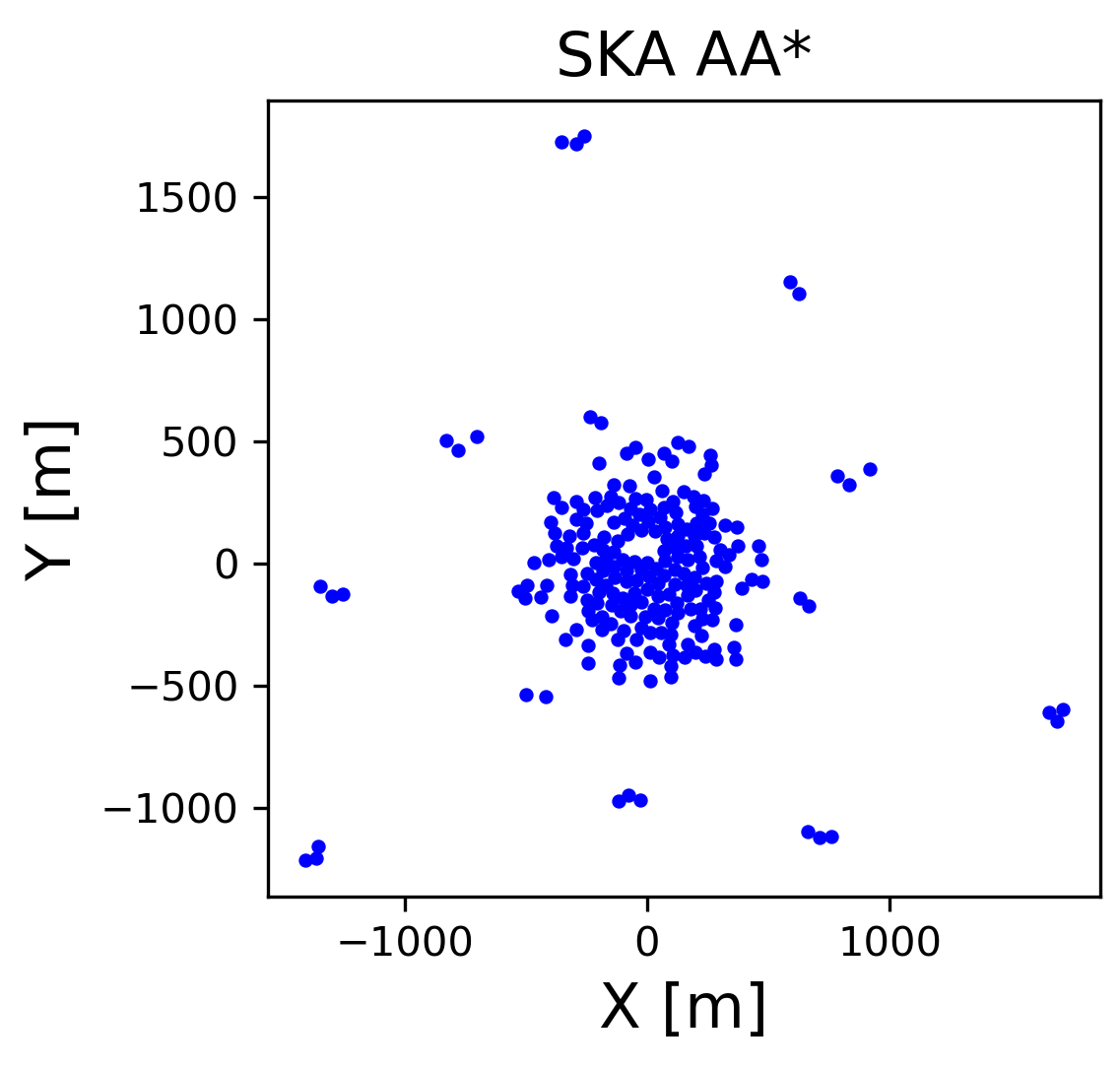}}
    \caption{Telescope configuration utilized in the simulation: SKA-Low AA* (with a 2 km central core).}
   
    \label{Fig_array}
\end{figure*}

In this work, we focus primarily on the upcoming SKA-Low telescope. The Square Kilometre Array (SKA)\footnote{\url{https://www.skao.int/en/explore/telescopes/ska-low}} is a next-generation, highly sensitive radio telescope designed to detect low-frequency radio signals, such as the 21-cm signal, making it a powerful tool for probing the Epoch of Reionization (EoR) \citep{2015Koopmans}. The SKA-Low array design includes 512 stations (referred to as AA4), each with a 35-meter diameter, comprising 256 antennas per station. Approximately 50 \% of these stations will be concentrated within a 1 km diameter central core, while the remaining stations are distributed along three spiral arms, arranged in clusters of 6 stations with logarithmic spacing. The configuration allows for a maximum baseline of about 73.4 km. In this study, we focus on the $ \rm AA^{\ast}$ SKA-Low layout, which consists of 307 stations with a maximum baseline of 73.4 km. From this layout, we selected all stations located within a 2 km radius (equivalent to a maximum baseline of 2000 meters from the central station), resulting in a subset of 231 stations, array shown in Fig.~\ref{Fig_array}. 
 
\begin{table}
\centering
\begin{tabular}{cc}
\hline
Parameters & SKA ($ \rm AA^{\ast}$)  \\\hline
Central Frequency    & 142 MHz (z$\sim$9)  \\     

Bandwidth   & 8 MHz        \\  
Number of frequency channels    & 64   \\
Image Field of view & 4$^\circ$          \\      
Number of array elements ($N_{\mathrm{a}}$)  & 231  \\     
Maximum baseline (m)& $\sim$2000  \\

Effective collective area ($A_{\mathrm{eff}}$)  &  $\sim$ 962 $m^{2}$   \\      
Core area of an array ($A_{\mathrm{core}}$)  &  $\sim$ 12.57 $km^{2}$   \\ \hline    
                                                                   
\end{tabular}
\caption{Parameter values used to conduct these synthetic simulations.}
\label{tab:sim_parms}
\end{table}

\section{Power Spectrum}
\label{sec:ps}
A key scientific objective of both ongoing and upcoming radio interferometric arrays is to detect and characterize fluctuations in the brightness temperature of the redshifted 21cm signal from the EoR. Various statistical tools are available in interferometric experiments to estimate the 21-cm brightness temperature fluctuations through the power spectrum (PS), either directly from the measured visibilities using the delay spectrum technique \textbf{\citep{2012Parsons, 2019Morales}}, or via image-based methods that involve gridding, calibration, and deconvolution to reconstruct the full 2D or 3D power spectrum \citep{lofar2, Patil2017}. In this study, we compute the theoretical power spectrum (PS) from the image plane and the observed PS from simulated visibility data using the SKA-Low layout, as illustrated in Fig.~\ref{Fig1}. 

\subsection{Theoretical Power Spectrum}
The fluctuation of the brightness temperature $ \rm \delta T_{b} $ for the 21 cm signal can be defined as \citep{2013Mellema}:

\begin{equation}
\delta T_{b}(\nu) \sim 27x_{HI}(1+\delta_{m})\Bigg(\frac{H}{\frac{dv_{r}}{dr}+ H}\Bigg)\Bigg(1-\frac{T_{\gamma}}{T_{S}} \Bigg) \times \Bigg(\frac{1+z}{10} \frac{0.15}{\Omega_{m} h^{2}} \Bigg)^{1/2} \Bigg(\frac{\Omega_{m} h^{2}}{0.023} \Bigg) \Bigg(\frac{\Omega_{b} h}{0.031} \Bigg) \, [\text{mK}]
    \label{Bright_temp}
\end{equation}

where, $x_{\mathrm{HI}}$ denotes the neutral hydrogen fraction, $\Omega_{\mathrm{m}}$ represents the matter overdensity, and $\rm H$ refers to the Hubble parameter. Additionally, $\rm dv_{r}/dr$ signifies the local gradient of gas velocity along the line of sight, while $T_{\mathrm{S}}$ and $T_{\mathrm{\gamma}}$ correspond to the spin temperature of the intergalactic medium (IGM) and the temperature of the cosmic microwave background (CMB), respectively.

We can define the dimensionless 21cm power spectrum as :

\begin{equation}
   \Delta^{2} (k) = \frac{k^{3}}{2\pi^{2}} P(\mathbf{k})
   \label{3d_ps}
\end{equation}

 $P(\mathbf{k})$ can define as 
 \begin{equation}
      \rm < \delta T_{b}(k) \delta T_{b}(k')> = (2 \pi)^{3}\delta(k+k')P(\mathbf{k})
 \end{equation}
where $\delta T_{\mathrm{b}}(k)$ is Fourier conjugate of  $\delta T_{\mathrm{b}}(\nu)$. 
\subsection{Observational Power Spectrum}
The interferometric visibility is defined as the correlation between the signals received by a pair of antennas, which is given by \citep{Thyagarajan2015a, Taylor1999} :

\begin{equation}
    V ({\mathbf{U}},\nu) = \iint A(\hat{\mathbf{s}},\nu) B(\nu) I(\hat{\mathbf{s}},\nu) e^{-i2\pi \nu \mathbf{U}.\hat{\mathbf{s}}} d\Omega,
    \label{vis_eq}
\end{equation}

Here, ${\mathbf{U}}$ represents the baseline vector, while $I(\hat{\mathbf{s}},\nu)$ and $A(\hat{\mathbf{s}},\nu)$ correspond to the specific intensity and the antenna beam pattern, respectively, both as functions of frequency ($\nu$). The term $B(\nu)$ represents the instrumental bandpass response. For this simulated observation, we assume an isotropic, frequency-independent antenna beam pattern, resulting in a flat bandpass response with $B(\nu)$ = 1. The unit vector is defined as $\hat{\mathbf{s}} \equiv (l,m,n)$, where $l$, $m$, and $n$ are the direction cosines pointing towards the east, north, and zenith, respectively, with $n = \sqrt{1-l^{2}-m^{2}}$. The solid angle element is given by $ d\Omega = \frac{dl dm}{\sqrt{1-l^{2}-m^{2}}}$. In this study, $ A(\hat{\mathbf{s}},\nu)$ is assumed to be 1, meaning the primary beam effect is not taken into account. 

The inverse Fourier transform of $V ({\mathbf{U}},\nu)$ along the frequency axis converts the visibility into the delay domain, denoted as $\tau$, resulting in $V ({\mathbf{U}},\tau)$. Based on this approach, the cylindrical power spectrum, as described in \citep{Thyagarajan2015a}, is expressed as follows:

\begin{equation}
    P(\mathbf{k}_{\perp}, k_{\parallel}) = \Big(\frac{\lambda^{2}}{2k_{B}}\Big)^{2}    \Big(\frac{X^{2}Y}{\Omega B}\Big)  |V (\mathbf{U},\tau)|^{2}, 
    \label{2dps_eq}
\end{equation}

where $\lambda$ represents the wavelength corresponding to the central frequency, $k_{\mathrm{B}}$ is the Boltzmann constant, $\Omega$ denotes the primary beam response, and $B$ is the bandwidth. The factors $X$ and $Y$ convert angular and frequency measurements into the transverse co-moving distance $D(z)$ and the co-moving depth along the line of sight, respectively \citep{Thyagarajan2015a}. The term $\mathbf{k}{\perp}$ corresponds to the Fourier modes perpendicular to the line of sight, while $k{\parallel}$ represents the modes along the line of sight, defined as follows:
 \begin{equation*}
 \mathbf{k}_{\perp} = \frac{2\pi |\mathbf{U}|}{D(z)} \hspace{20pt} \& \hspace{20pt}
  k_{\parallel} \mathbf{\approx} \frac{2\pi \tau \nu_{21} H_{0} E(z)}{c(1+z)^{2}}
 \end{equation*}
Here, $\nu_{21}$ represents the rest-frame frequency of the 21 cm spin-flip transition of \hi, while $z$ denotes the redshift corresponding to the observing frequency. The Hubble parameter is given by $H_{\mathrm{0}}$, and $E(z) \equiv [\Omega_{\mathrm{\mathrm{M}}}(1+\textit{z})^{3} + \Omega_{\mathrm{\Lambda}}]^{1/2}$. The parameters $\Omega_{\mathrm{M}}$ and $\Omega_{\mathrm{\Lambda}}$ correspond to the matter and dark energy densities, respectively \citep{1999Hogg}. 

The 1D power spectrum is derived from the 2D power spectrum by performing a spherical average of $P(\mathbf{k}{\perp}, k{\parallel})$ and is given by:
\begin{equation}
   \Delta^{2} (k) = \frac{k^{3}}{2\pi^{2}} <P(\mathbf{k})>_{k}
   \label{3d_ps_ii}
\end{equation}
where, $k = \sqrt{k_{\perp}^{2} + k_{\parallel}^{2}}$.

\begin{figure*}
    \centering
    {\includegraphics[width=0.8\textwidth]{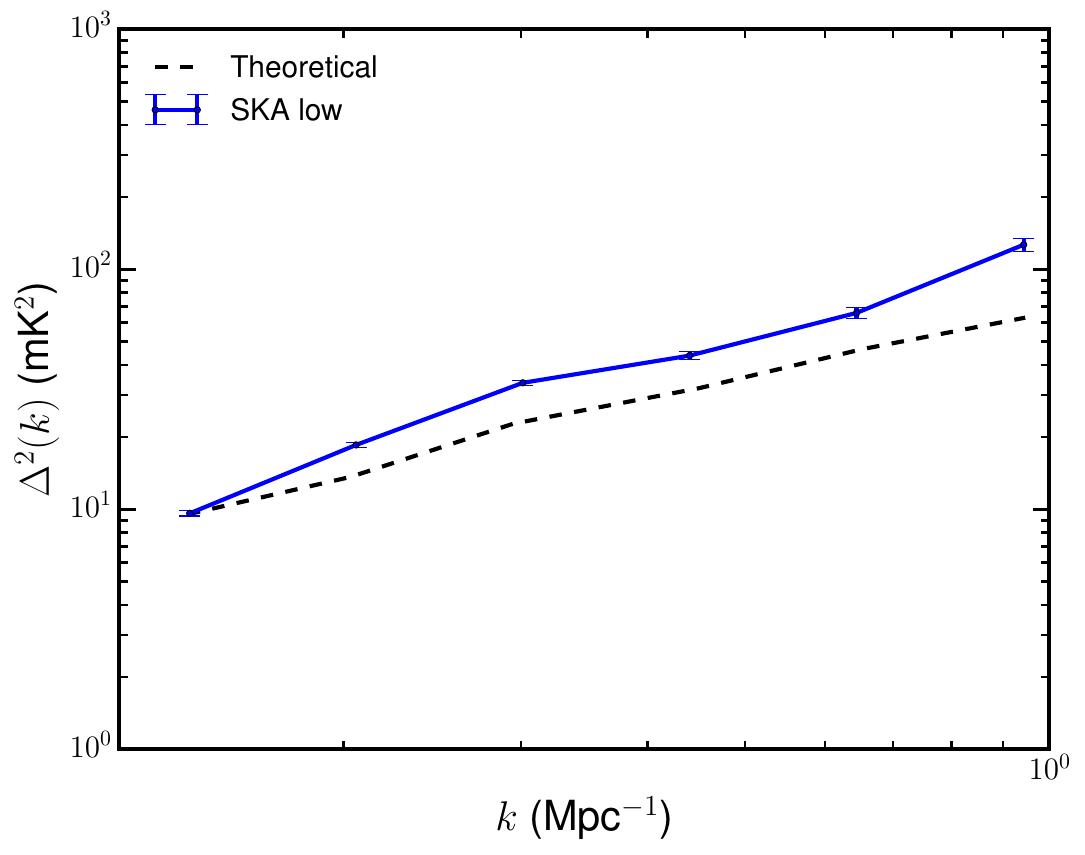}}
    \caption{Shows a comparison between the theoretical spherical power spectrum (PS) and the simulated observed PS for the same sky model. The theoretical PS for the signal models is represented by a black dashed line, while the simulated observed PS for the SKA-Low array configurations is shown as solid blue lines.}
   
    \label{Fig1}
\end{figure*}

\section{Emulating {\hi} 21cm Power Spectrum}
\label{sec:emu ps}
The objective of developing ANN-based emulators for this work is to study the power spectrum, as these emulators can generate efficient and reliable EoR models. 
ANNs are a class of machine learning models inspired by the neural architecture of the human brain. They consist of multiple layers of interconnected computational units, referred to as neurons, which process and propagate information through the network. Each neuron applies a transformation to its input via a mathematical function known as an activation function, and passes the result to subsequent layers. Hidden layers those situated between the input and output layers facilitate the learning of complex and non-linear relationships within the data. Commonly used activation functions such as the Rectified Linear Unit (ReLU) and the Exponential Linear Unit (ELU) introduce the necessary non-linearity for the network to approximate highly intricate mappings between inputs and outputs.

\begin{figure*}
    \centering
    {\includegraphics[width=0.45\textwidth]{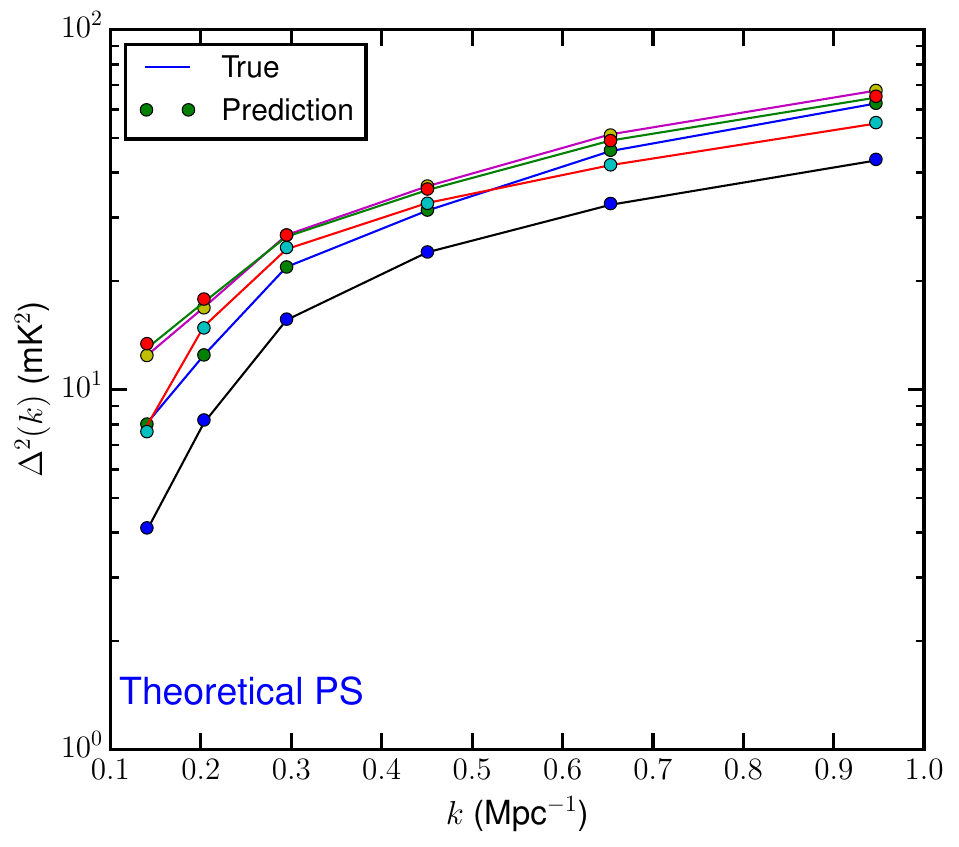}}
    {\includegraphics[width=0.45\textwidth]{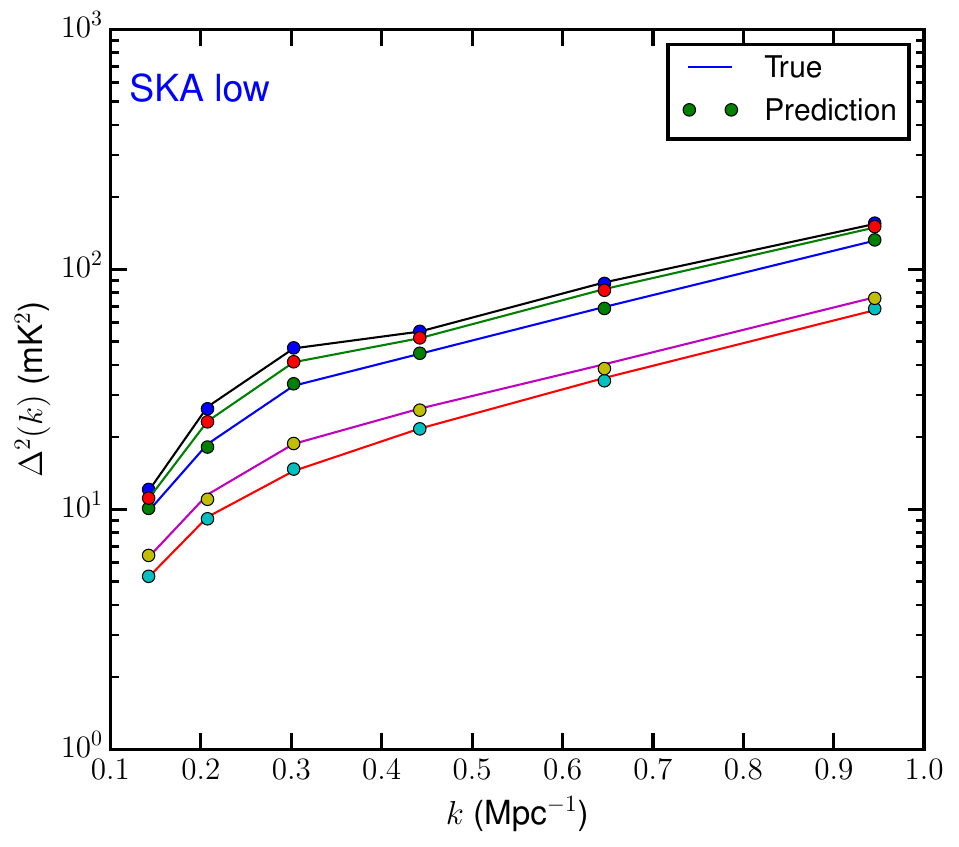}}
    \caption{The figure shows a comparison between the simulated true power spectrum (solid lines) and the emulated power spectrum by the ANN (dots). The \textbf{left panel} shows emulator predictions trained on theoretical power spectra, while the \textbf{right panel} presents predictions based on observed power spectra from SKA-Low.}
   
    \label{Fig2}
\end{figure*}

These models can then be utilized as substitutes for computationally intensive simulations in Bayesian parameter inference. In addition to this, a non-parametric feature of ANN emulators is the ability to replicate various signal features without relying on their specific parametric characteristics, as they are solely dependent on the training data sets. This capability allows us to utilize the features directly without the need to remove or apply any cleaning algorithms, thereby enabling the inference of associated label parameters. In this study, we are developing an emulator by training it on a set of mock observed power spectra generated from simulated observations with the SKA-Low interferometric array. We also train the emulator on the theoretical power spectrum, that is, in the absence of both instrumental effects and array configuration influences. We use the ANN model to build these emulators. To develop ANN architecture, we are using the Python-based deep learning Keras API, and standard sci-kit learn \citep{2011Pedregosa}. To achieve significant accurate PS from the emulator for test set of EoR parameter we have to train the emulator with the sufficient number of the training datasets. 
To construct an optimal training data set, we sample the parameter space with Latin Hypercube Sampling. Latin Hypercube sampling method samples the multi-dimensional parameter space such that no two parameters share the same value in the parameter space, providing an all-unique set of parameters.
We generated 300 unique combinations of astrophysical parameters using Latin Hypercube Sampling from the defined parameter ranges to construct the training and test datasets. These sampled parameter sets were used to simulate lightcones, as detailed in Section \ref{sec: HI_sim}, and subsequently to compute the observed power spectra using the 21cm end-to-end (E2E) pipeline. To develop construct training and test datasets for the theoretical PS, we use 21cmFast. The parameters range we follow to construct the training datasets are $R_{\mathrm{mfp}}$ = (10 Mpc, 60 Mpc ), $log(T_{\mathrm{vir}})$ = (4.5 K, 6.0 K ), and $\zeta $ = (10, 60). We construct training and test data sets of observed/theoretical power spectrum ($\Delta^{2}(k)$) by following these sample parameters for 6 different $k$ modes.  We use 270 PS data sets to train and validate the ANN model and 30 data sets to test the ANN emulator. We constructed the network architecture by following Python-based Keras’ Sequential API. The network consists of an input layer consisting 3 neurons matching with the training 3 EoR key parameter ($R_{\mathrm{mfp}}$, $\log(T_{\mathrm{vir}})$, $\zeta$) and two hidden layers with 28 and 14 neurons, respectively, each activated by the ’elu’ activation function. The number of neurons in the hidden layers is determined empirically through trial and error, based on the configuration that yields the best performance in terms of accuracy and robustness. The output layer has 6 neurons to predict the observed PS. To ensure effective model training and convergence, the input PS values are standardized using the \texttt{StandardScaler} function, which removes the mean and scales the features to unit variance. The corresponding EoR parameters are normalized to the [0, 1] range using \texttt{MinMaxScaler}. Both scalers are implemented using the \texttt{sklearn} library. This preprocessing step is essential for bringing all input features to a comparable scale, thereby enhancing training stability and learning efficiency. We trained this ANN model using different PS observed by SKA-Low interferometric arrays, to create emulator. Fig.~\ref{Fig2} presents a comparison between the true test set 21-cm PS, shown as solid lines, and the emulated PS by the ANN, represented by dots. For this comparison, we randomly selected five test sets of power spectra and compared them with the predicted PS from the emulators in each case.

\section{Bayesian Inference of EoR Parameters}
\label{sec:Bayesian_Inference}
To constrain the EoR parameters and associated errors in the inferred parameter values, the Bayesian approach is one of the most popular methods. In the Bayesian approach, the posterior distribution of the parameters can be defined using Bayes' theorem: 
\begin{equation}
    p(\theta|D,M) = \frac{p(D|\theta,M)  \Pi(\theta|M)}{p(D|M)}
\end{equation}

where $\theta$ represents the parameters, D is data, and M represents a model. The evidence $p(D|M)$ serves as a constant normalization factor for a given model. The posterior distribution is determined solely by the product of the likelihood function $L = p(D|\theta, M)$ and the prior distribution of the parameters $\theta$, denoted as $\Pi(\theta|M)$. For this study, we considered a multivariate Gaussian likelihood which can expressed as:

\begin{equation}
    ln L = -\frac{1}{2}[\Vec{d}_{ref} -\mu]^{T}[\sigma^{2}]^{-1}[\Vec{d}_{ref} -\mu] -\frac{1}{2}ln(2\pi\, \det\{\sigma^{2}\})
\end{equation}

where $\Vec{d}_{\mathrm{ref}}$ represents reference data, $\mu$ represents the model observable corresponding to the observable parameters and $\sigma^{2}$ is the error covariance associated with $\Vec{d}_{\mathrm{ref}}$.

We design our pipeline to extract the most likely EoR parameter values using MCMC for the given log-likelihood. We used the CosmoHammer \citep{cosmohammer} Python package, utilizing an affine-invariant \cite{affine2010} Markov Chain Monte Carlo (MCMC) ensemble sampler, to conduct model parameter estimation. We have tested the performance of the pipeline using both the theoretical power spectrum and the simulated observed power spectrum. Further details are discussed in Section~\ref{sec:result}.

\section{Error Covariances of Power Spectrum}
\label{sec:err_cov}
The error estimate in the observables is essential to predict the posterior of the parameters. The EoR 21cm signal has varoius contamination from the foreground, systematics, noise, calibration errors etc. To estimate the error covariance of the power spectrum (PS), we assume that the foregrounds have been perfectly modeled and entirely removed from the observed signal. This perfect modelling assume of the sky and instrument to establish a baseline for evaluating performance under idealized conditions. This allows us to isolate the intrinsic capabilities and limitations of the methodology without complications from model inaccuracies. In the following section, we introduce modelling imperfections such as gain calibration and sky position errors to assess the robustness of the method and quantify their impact on power spectrum estimation. This comparison highlights the potential biases that arise when instrumental and modelling errors are neglected, motivating future work on mitigation strategies. Under the perfect modelling assumption, the remaining observed signal consists of the EoR 21cm signal and Gaussian thermal noise. Hence, in this analysis, the total covariance ($\sigma_{\mathrm{t}}^2$) is defined as the sum of the sample variance ($\sigma_{\mathrm{SV}}^2$) and the thermal noise variance ($\sigma_{\mathrm{N}}^2$). 

\begin{equation}
    \sigma^{2}_{t}(i) = \sigma^{2}_{SV}(i) + \sigma^{2}_{N}(i) 
\end{equation}

In this analysis, we assume that the measurements at any two bins are mutually uncorrelated which simplifies the error computation. In this study we use variance instead of covariance to calculate error $\rm \sigma^{2}$. 
We estimated the sample variance for the bin average power spectrum $P(k_{\mathrm{i}}$) by following equation 

\begin{equation}
    \sigma^{2}_{SV}(P_{i}) = [\Delta^{2}(k_{i})]^{2}/N_{k_{i}}]
\end{equation}
where $N_{\mathrm{k_{i}}}$ denotes the number of measurement at each k-mode.
The thermal noise for given radio interferometric array can be estimated using the following equation \citep{McQuinn2006, Mellema2013}:

\begin{equation}
    \sigma_{N}(P_{i}) = {\frac{k^{3}}{2\pi^{2}\sqrt{N_{k_{i}}}}}\left(\frac{2T^{2}_{sky}}{Bt_{obs}}\frac{D^2(z)\Delta D \Omega_{FoV}}{n_{p}}\left(\frac{A_{eff}A_{core}}{A^{2}_{coll}}\right)\right)
\end{equation}

where $\rm A_{\mathrm{eff}}$, $\rm A_{\mathrm{coll}}$, and $\rm A_{\mathrm{core}}$ denote the effective area, collecting area, and core area, respectively, of the specified radio interferometric array. The symbol $\rm t_{\mathrm{obs}}$ signifies the total observation time in hours. Additionally, $\rm D$ and $\rm \Delta D$ denote the comoving distance to the redshifts where the center is located and the comoving distance corresponding to a bandwidth $\rm B$ at that comoving distance.  $ \rm T_{sky}$ denotes the sky temperature calculated using $\mathrm{T_{\mathrm{sky}} \sim 180\Big(\frac{\nu}{180 MHz}\Big)^{-2.6}}$K \citep{Furlanetto2006}, symbol $\Omega_{\mathrm{FoV}}$ denotes field of view symbol, $n_{\mathrm{p}}$ represents the number of polarization and $N_{\mathrm{k_{i}}}$ denotes the number of measurement at each k-mode.

\section{Result}
\label{sec:result}
This section presents the simulation results and discusses their implications for real observations. Detailed outcomes for each case are provided in the following subsections.

\subsection{Perfect Observing Condition}
\begin{figure*}
    \centering
    {\includegraphics[width=0.9\textwidth]{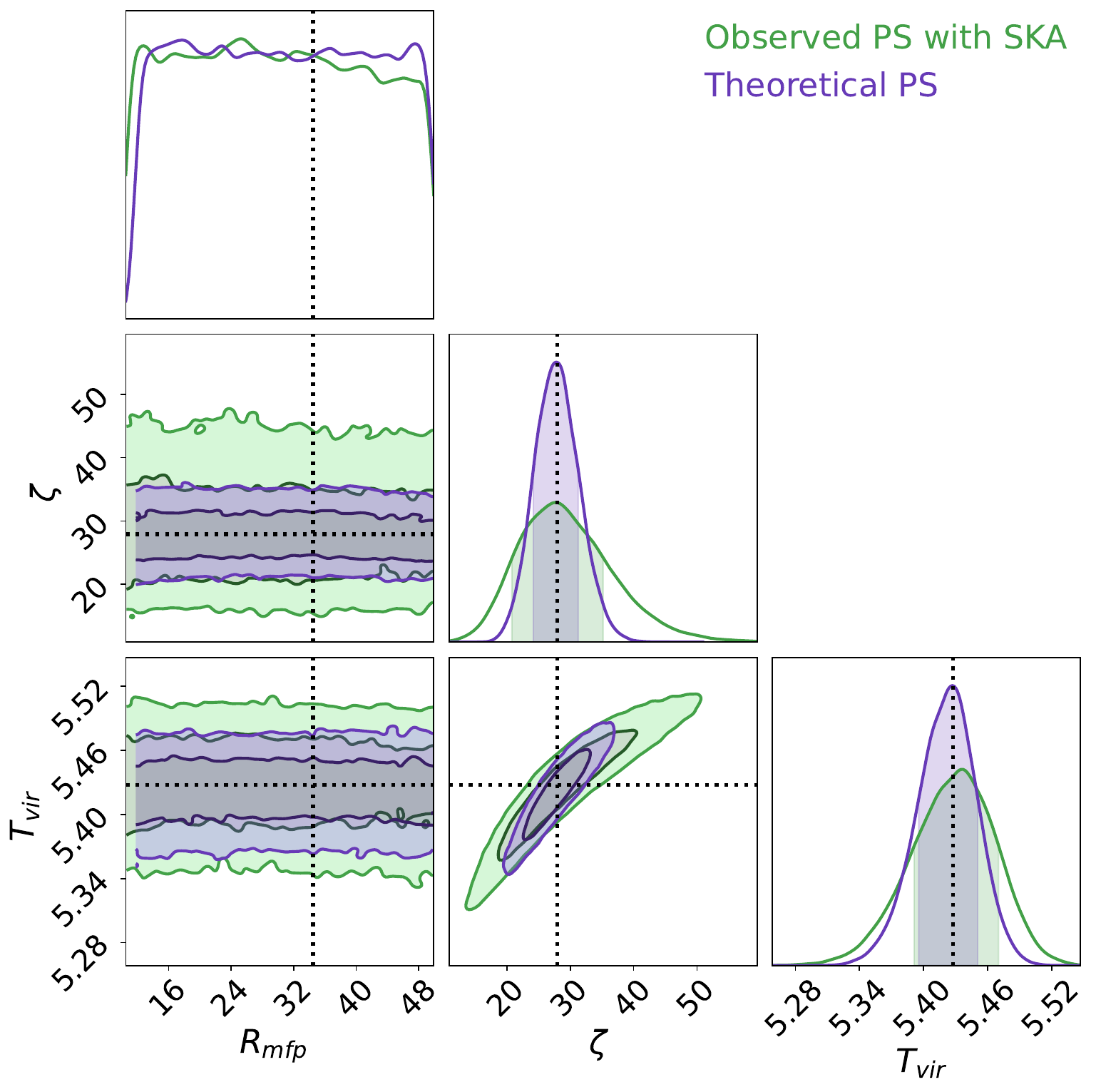}}
    \caption{Depicts the posterior distribution of model parameters obtained through power spectrum analysis, comparing the theoretical power spectrum with the observed power spectrum from SKA-Low. The enclosed areas between the inner and outer contours signify the 1$\sigma$ and 2$\sigma$ confidence levels, respectively. }
    \label{Fig3a}
\end{figure*}

In this study, we infer astrophysical parameters from the power spectra (PS) obtained using the SKA-Low interferometric array. We assumed that foregrounds were perfectly removed from the observed PS and that there were no calibration or positional errors, as shown in Fig.~\ref{Fig1}. To achieve this, we simulated observations using the 21cm E2E pipeline with only input {\hi} lightcone maps and calculated the PS from the resulting observed {\hi} visibility. To derive the astrophysical parameters, we employed an emulator-based Markov Chain Monte Carlo (MCMC) pipeline. The advantage of employing this emulator-based MCMC pipeline lies in its ability to reduce computational time significantly. Furthermore, since the emulators are trained on observed power spectra that inherently incorporate the effects of the telescope layout, they enable direct parameter inference without requiring separate corrections for these effects. The astrophysical parameters inferred using the SKA-Low interferometric array are presented in Fig.~\ref{Fig3a}. In this case, we found that the two parameters, $\zeta$ and ${T_{\mathrm{vir}}}$, closely matched the actual values. However, the third parameter, $R_{\mathrm{mfp}}$, was not correctly constrained. To investigate this issue, we examined whether the lack of constraint could be due to the effects introduced by the telescope layout. Additionally, we compared this result with the inferred astrophysical parameters derived from theoretical power spectra. A similar trend was observed in this case as well. Specifically, $\zeta$ and ${T_{\mathrm{vir}}}$ closely matched the actual values, as shown in Fig.~\ref{Fig3a}. However, $R_{\mathrm{mfp}}$ remained unconstrained, which may be attributed to the degeneracy of the $R_{\mathrm{mfp}}$ parameter. To further assess the robustness of this pipeline, we inferred parameters for two additional power spectra generated using astrophysical parameters near the boundary values of those used in the training dataset. The inferred parameters for these power spectra are presented in Fig.~\ref{Fig_ps}, with set 2 on the left and set 3 on the right, as detailed in Appendix \ref{Robustness_check}. Similar to the initial case, the astrophysical parameters $\zeta$ and $\mathrm{T_{\mathrm{vir}}}$ were well-constrained and closely aligned with the true values in most instances, as depicted in Fig.~\ref{Fig_ps}. However, the parameter $R_{\mathrm{mfp}}$ remained poorly constrained across both sets.\ As noted in \cite{schmit2018emulation}, this degeneracy arises around redshift $z \simeq 9.0$, where a population of low-mass galaxies with brighter stellar populations can replicate the fiducial observation. These galaxies tend to ionize the IGM earlier, and as a result, combining observations across multiple redshifts and incorporating the evolution of the ionization process can help break this degeneracy. In future work, we plan to test this explicitly by including multi-redshift observations to assess the improvement in parameter recovery.

We also noticed that the power spectrum from set 1, whose astrophysical parameters are around the centre of the boundaries, generated more accurate conclusions. By contrast, the power spectra generated in close proximity to the boundary values of the astrophysical parameters yielded an inferred parameter mean that slightly differed from the actual value. For example, in Set 3, the true value of $ \rm \zeta$ is 57.42, while the inferred mean using the theoretical power spectrum is 50.8. Using the SKA power spectrum, the inferred mean is 59.8. Although these values differ from the true mean, they still lie within $1\sigma $ of the inferred uncertainty range (see Table~\ref{tab:inferred}). This mismatch could be due to the ANN emulator being trained on less training datasets. Potential future development of a more precise emulation could involve training it with a more extensive dataset that cover the parameter space in the effective manner.   

\subsection{Imperfect Observing Conditions}
Sensitive radio observations targeting the {\hi} signal from the CD and EoR are highly susceptible to contamination from various sources. Small errors in early data processing steps, such as calibration, can propagate into the final power spectrum (PS) estimates, potentially leading to misinterpretation of the results. In \citep{2022Mazumder}, the authors investigated the effects of gain calibration errors and sky model position errors on the recovery of the 21-cm power spectrum. In this study, we build upon that work by examining how such errors impact the recovery of astrophysical parameters. \cite{2022Mazumder} previously conducted a visual comparison of the power spectrum, analyzing both 2D and 1D residual PS in the presence of gain calibration and sky position errors. They also evaluated how the root mean square (RMS) of the residual image varies with different levels of gain calibration and position errors, comparing these values against the signal and thermal noise levels. Their findings showed that a gain calibration error as small as 0.01\% could result in a high image RMS in SKA-Low observations, potentially causing confusion or even masking faint cosmological signals. Likewise, a sky model position error exceeding 0.48 arcseconds was found to significantly elevate the image RMS, leading to similar challenges in signal detection. In our analysis, we extend this investigation by comparing the corrupted 1D power spectrum with the original and estimating the astrophysical parameters from the corrupted spectra in two scenarios: one with gain calibration errors and another with sky model position errors.

\subsubsection{Gain Calibration Errors}
Gain calibration errors were incorporated by applying complex gain perturbations to each antenna individually, as expressed in Equation~\ref{eq:measured_visibility}, following the methodology of \cite{2022Mazumder}. These per-antenna errors propagate through the visibilities since each visibility is formed from a pair of antennas and thus impact the entire $k$ space range. The visibility measured between a pair of antennas $i$ and $j$, denoted as $V_{\mathrm{ij}}^m(t)$, is given by:

\begin{equation}
    V_{ij}^m(t) = g_i(t)\, g_j^*(t)\, V_{ij}^t(t)
    \label{eq:measured_visibility}
\end{equation}

Here, $V_{\mathrm{ij}}^t(t)$ is the true visibility from the sky, while $g_{\mathrm{i}}(t) $and $g_{\mathrm{j}}(t)$ are the complex gains of antennas $i$ and $j$, respectively. Although gain calibration attempts to estimate and correct for these instrumental effects, it is typically imperfect, leaving behind residual errors that can propagate through the analysis and bias the recovered signal.

The complex gain for each antenna can be modeled as:

\begin{equation}
    g_i = (a_i + \delta a_i)\, \exp[-i(\phi_i + \delta \phi_i)]
    \label{eq:gain_model}
\end{equation}

where $a_\mathrm{i}$ is the nominal amplitude and $\phi_{\mathrm{i}}$ the phase (in radians), with $\delta a_{\mathrm{i}}$ and $\delta \phi_{\mathrm{i}}$ representing the residual errors in amplitude and phase, respectively. In an ideal scenario, $a_\mathrm{i} = 1 $ and $\phi_\mathrm{i} = 0 $, yielding a perfect gain of 1. However, with calibration imperfections, the gain simplifies to:

\begin{equation}
    g_i = (1 + \delta a_i)\, \exp[-i\, \delta \phi_i]
    \label{eq:residual_gain}
\end{equation}

\begin{figure*}
    \centering
    {\includegraphics[width=0.45\textwidth]{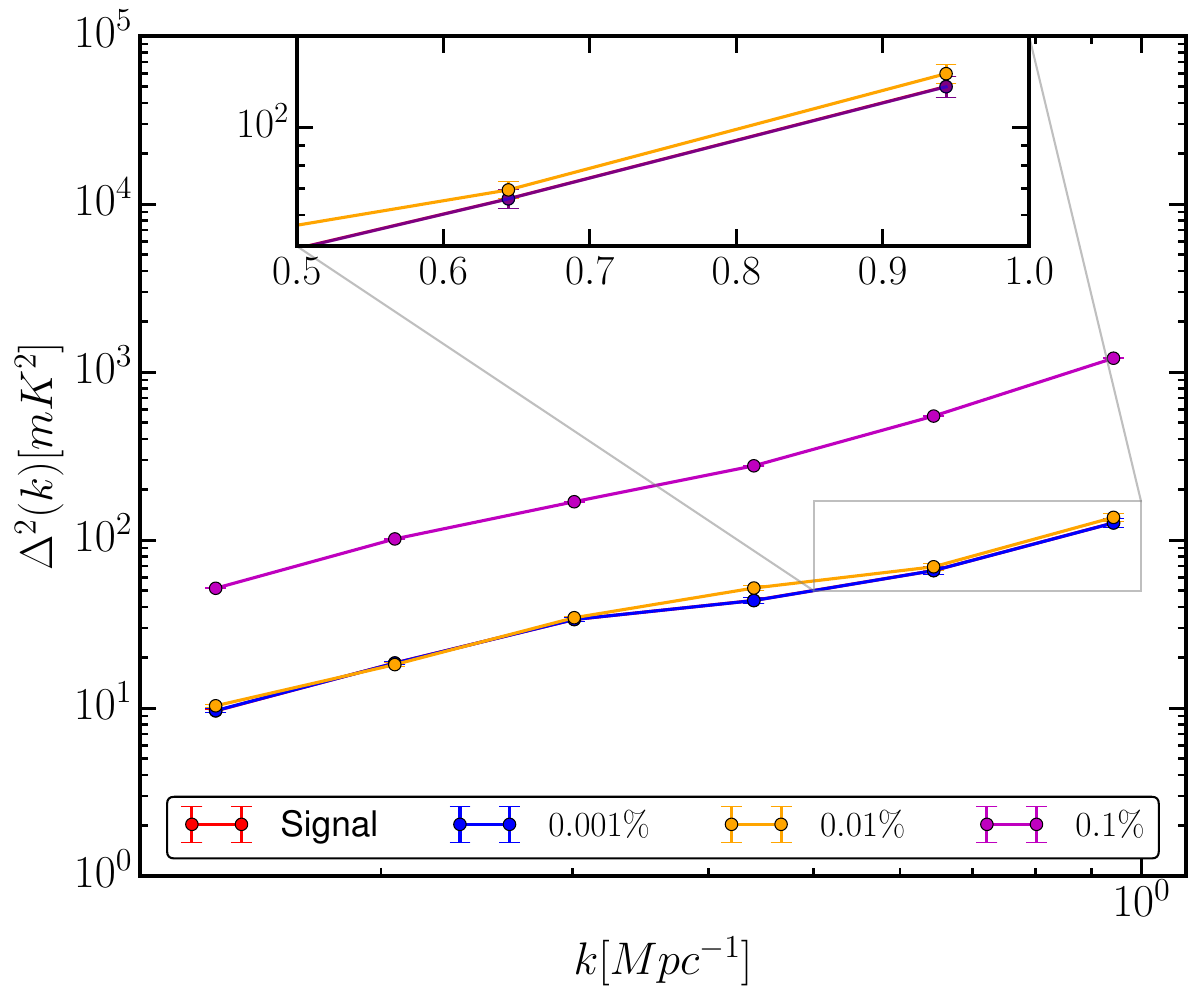}}
    {\includegraphics[width=0.45\textwidth]{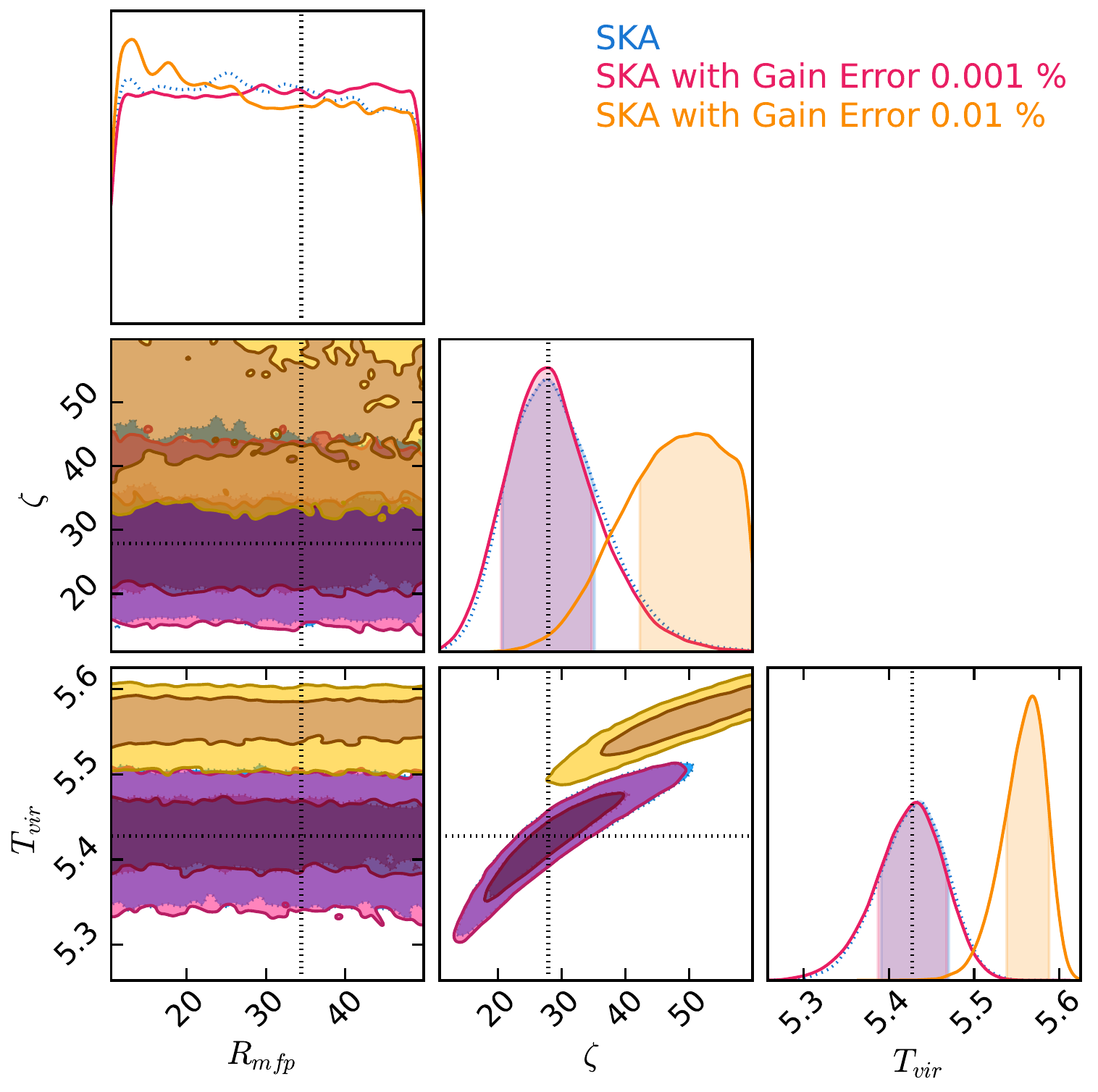}}
    \caption{\textbf{(Left)} Residual power spectra for gain calibration errors (0.001\%, 0.01\%, and 0.1\%) compared to the signal power for SKA-Low array layouts. Error bars represent 1$\sigma$ uncertainties, including sample variance and thermal noise. \textbf{(Right)} Posterior distributions of model parameters from the power spectrum analysis (blue), with gain calibration errors of 0.001\% (magenta) and 0.01\% (orange) for SKA-Low.The shaded regions represent the 1$\sigma$ and 2$\sigma$ confidence intervals.}
   
    \label{Fig3}
\end{figure*}

To assess the influence of gain calibration errors on the observed power spectrum (PS) and their subsequent impact on associated astrophysical parameters for SKA-Low, we introduced various gain calibration errors as detailed in \cite{2022Mazumder}. The resulting residual PS was visually compared with the actual PS, depicted in the left panel of Fig.~\ref{Fig3}. The parameter obtained from the PS analysis with SKA-Low, incorporating actual signal and PS for gain error residuals, is depicted in the right panel of Fig.~\ref{Fig3}. Upon examining the plotted residual PS spectrum, we observed that for gain calibration error of 0.001 \%, the residual PS overlaps with the true PS. For residual PS with a gain calibration error of 0.01\%, we observe overlapping at lower k modes but with minor deviation at higher k modes. However, when deriving astrophysical parameters from the residual power spectrum, we found that for a gain calibration error of 0.001\%, the derived parameters match those derived from the true PS. In contrast, for PS with the gain calibration error of 0.01\%, the derived parameters exhibit significant differences from the actual values. This indicates that astrophysical parameters are highly sensitive to gain calibration errors, and even minor deviations in the PS due to gain calibration errors can severely impact the estimation of parameters, introducing bias. The sensitivity of derived parameters to gain calibration errors clearly suggests that the gain calibration error threshold should be much lower than 0.01\%, ideally closer to 0.001\%. If the gain calibration error exceeds 0.001\%, it is advisable to model them to prevent bias in the derived parameter values.

\begin{figure*}
    \centering
     {\includegraphics[width=0.45\textwidth]{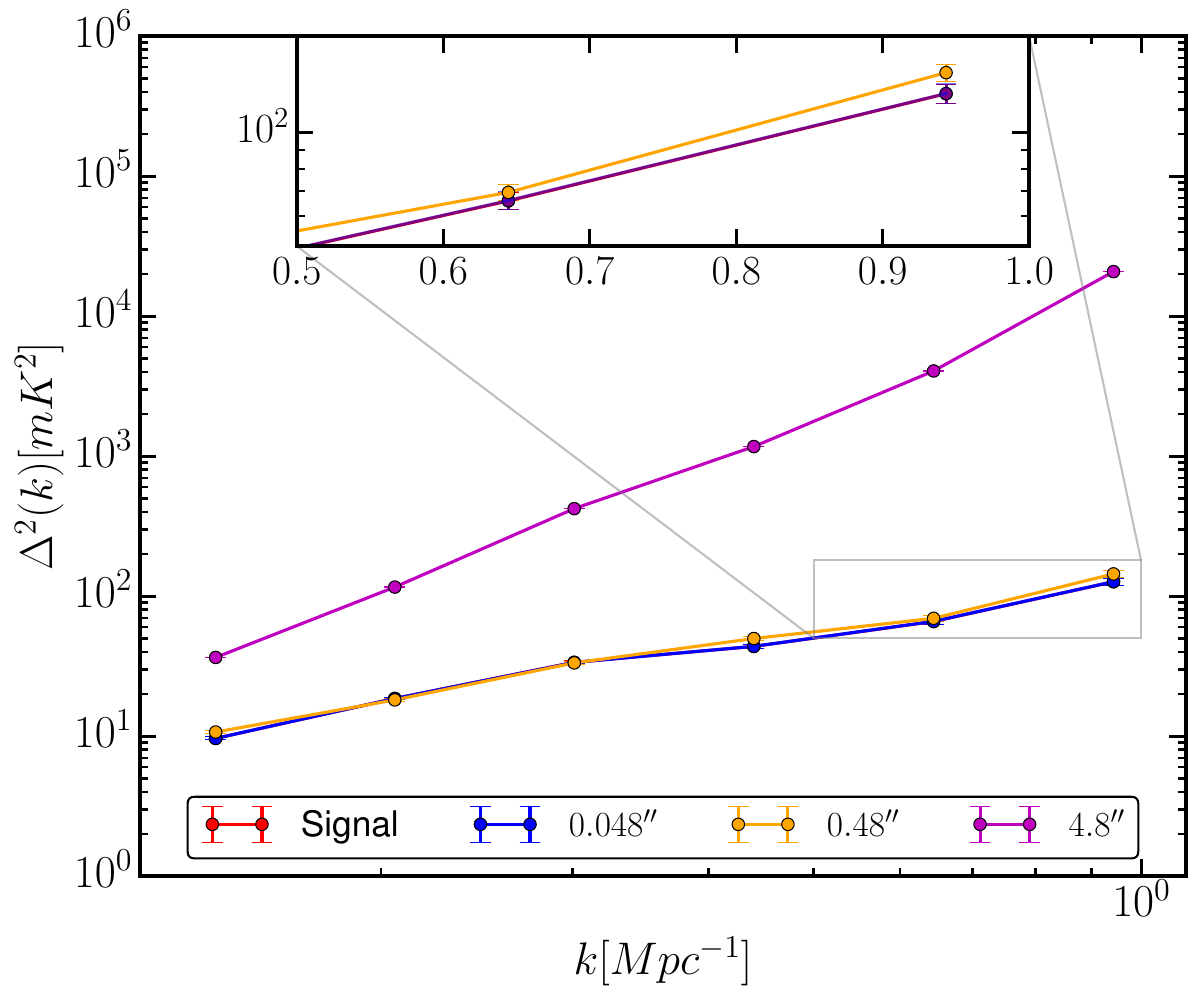}}
    {\includegraphics[width=0.45\textwidth]{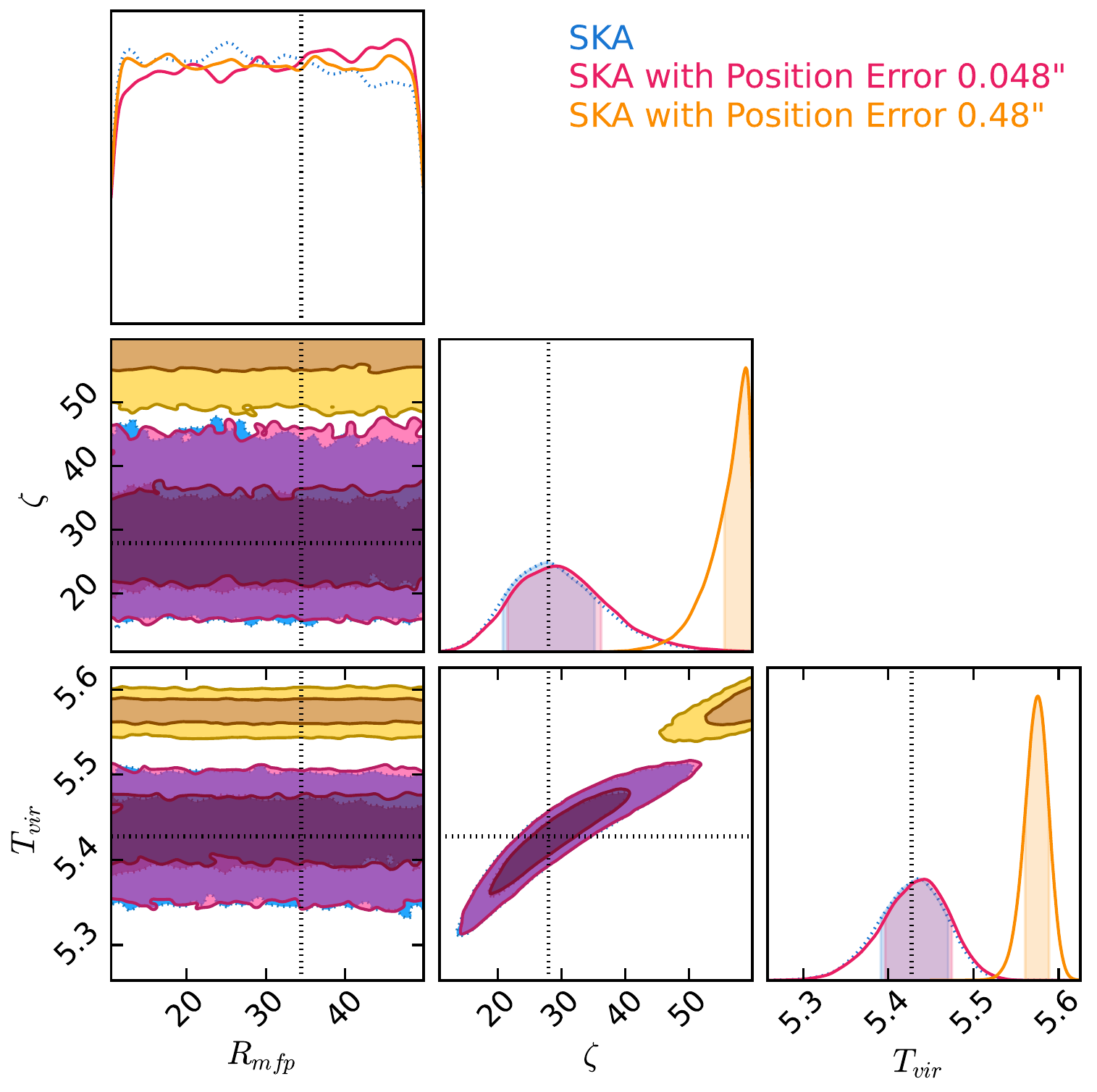}}
    \caption{\textbf{(Left)} Residual power spectra for position errors (0.048", 0.48", and 4.8") for SKA-Low. The error bars are 1$\sigma$ uncertainties for the k-bins including sample variance and thermal noise. \textbf{(Right)} Posterior distributions of model parameters from residual power spectra for position errors of 0.048" (magenta) and 0.48" (orange), with SKA-Low. The shaded regions represent the 1$\sigma$ and 2$\sigma$ confidence intervals.}
   
    \label{Fig4}
\end{figure*}

\subsubsection{Sky Model Position Error} 
To evaluate the impact of sky model position errors in the sky model, simulations were conducted using inaccurate sky models as described in \cite{2022Mazumder}. Sky model position errors were modeled with zero-mean Gaussian distributions and varying standard deviations (up to 0.048", 0.48", and 4.8") and applied to the right ascension (RA) of sources, resulting in a new catalog with positional inaccuracies. Residuals were then obtained following equation 4 from \cite{2022Mazumder}, by subtracting the corrupted sky from the true sky, and these residuals were used to analyze the effects on the PS. Unlike gain calibration errors which arise from instrumental calibration inaccuracies sky model position errors are rooted in prior assumptions about the sky and do not depend on the calibration solution itself. However, they can still introduce significant foreground residuals that mimic spectral structure, making them especially relevant for 21cm cosmology. This distinction is important, as position errors affect subtraction fidelity rather than the correction of instrumental response, and thus represent a separate and critical source of systematic error.

Similar to gain calibration errors, we evaluated the impact of sky model position errors on the observed power spectrum (PS) and their effect on the inferred astrophysical parameters for SKA-Low. Different sky model position errors, as described above, were introduced, and the resulting residual PS was compared with the true PS, as shown in the left panel of Fig.~\ref{Fig4}. The parameters obtained from the PS, which incorporates actual signals and position error residuals, are depicted in the right panel of Fig.~\ref{Fig4}. Upon examining the plotted residual PS spectrum, we found that for a sky model position error of 0.048", the residual PS overlaps with the true PS. At lower k modes, the residual PS with this position error overlaps well with the true PS, but at higher k modes, there is a minor deviation. This occurs because real-space scales are more affected by arcsecond-level errors at higher k values. However, since we used the entire k range to derive astrophysical parameters from the residual power spectrum, we observed that for a position error of 0.048", the derived parameters closely match those obtained from the true PS. Similarly, for a sky model position error of 0.48", the derived parameters exhibit substantial deviations from the true values. This indicates that astrophysical parameters are highly sensitive to sky model position errors, and even small deviations in the PS due to these errors can severely impact parameter estimation, introducing bias. The sensitivity of derived parameters to sky model position errors, similar to gain calibration errors, suggests that the acceptable sky model position error threshold should be much lower than 0.48", ideally around 0.048". If the sky model position error exceeds 0.048", modeling the errors becomes crucial to prevent bias in the derived parameter values.


\section{Summary and Discussion}
This study presents the development of an ANN and Bayesian-based framework for inferring astrophysical parameters from observed {\hi} power spectra (PS). We used the 21cm E2E pipeline to generate the mock observed power spectrum (PS) for SKA-Low core array configurations. However, generating a large number of modeled observable signals across a multi-dimensional parameter space for Bayesian inference is computationally expensive, particularly when using simulated observations. To address this, we adopted a formalism utilized by several groups, employing emulators for modelling the EoR signal instead of actual simulations.

For constructing the training datasets of observed PS, EoR 21cm signals were generated using the semi-numerical simulation 21cmFAST, with key EoR parameters ($ \rm R_{\mathrm{mfp}}$, $T_{\mathrm{vir}}$, and $\zeta$) described in section \ref{sec: HI_sim}. These signals served as sky signals for mock observations using the 21cm E2E pipeline with interferometric arrays like SKA Low, allowing us to construct sets of PS under different astrophysical conditions. These simulated observed PS were then used to train emulator, facilitating Bayesian inference of the corresponding astrophysical parameters. A key advantage of our ANN-based emulators, beyond conserving computational resources, is their ability to perform direct parameter inference using observed power spectrum (PS) data while inherently accounting for telescope layout effects. In radio interferometric observations, limited and uneven uv-coverage introduces deviations from the theoretical PS through sampling artifacts, resolution loss, and convolution with the uv-sampling window. Traditionally, inferring astrophysical parameters from the observed PS requires applying separate corrections to account for these instrumental and sampling effects before comparing with theoretical models. In our approach, however, the ANN models are trained directly on observed-like PS data that already include all such instrumental and sampling-induced distortions. As a result, the emulator learns to incorporate these effects during training, allowing for parameter inference directly from the observed PS without requiring additional correction steps. This contrasts with methods relying on theoretical PS models, where observed data must first be adjusted to match idealized conditions to enable accurate parameter estimation.

In the first case of the study, we perform the inference using the developed ANN and Bayesian-based pipeline assuming perfect observing conditions where we assume the foreground had been perfectly removing only signal and layout effect is there in the observed PS. We infer the astrophysical EoR parameters for the three sets for the signals constructed using the true astrophysical parameters, as listed in Tab.~\ref{tab:inferred}. Both sample variance and thermal noise (calculated for 1000 hours of observation with the given interferometer at redshift $z$ = 9) are incorporated in the inference. The detailed results are listed in Tab.~\ref{tab:inferred}. Additionally, we have inferred astrophysical parameters from theoretical power spectra (PS) and compared these findings with parameters inferred from observed PS. In all three cases, the inferred mean values of the parameters $\rm T_{\mathrm{vir}}$ and $\rm \zeta$ showed excellent agreement with the true input values, with the true values consistently falling within the 1$\sigma$ confidence intervals of the inferred posteriors (see Tab.~\ref{tab:inferred}). However, the parameter $\rm R_{\mathrm{mfp}}$ was often poorly constrained, indicating a high level of degeneracy.
In the second case of the study, we assumed that the foregrounds were not perfectly removed, leaving some residual foreground due to  gain calibration or sky model position errors. This study aimed to determine the threshold level of gain calibration or sky model position error for SKA-Low. In a previous study by \cite{2022Mazumder}, 2D and 1D power spectra with gain and position errors were visually examined. They also calculated the root mean square (RMS) in the residual image across different percentage errors in gain calibration and sky model position, comparing it with signal levels and thermal noise. They found the gain calibration error threshold to be 0.01 \% and the sky model position error threshold to be 0.48".

In this study, we delve deeper and systematically analyze these effects by performing inference on the observed power spectra upcoming SKA-Low. This allows us to understand the impact of various gain calibration and sky model position errors on astrophysical parameters. The key findings noted from this study:

\begin{itemize}
    \item Our analysis showed that the calibration error threshold for SKA-Low is 0.001\%. Exceeding this threshold results in significant deviations in the inferred astrophysical parameters from their actual values, severely affecting the astrophysical processes. An equivalent calibration threshold was identified by \cite{2016Barry}.
    
    \item The position error threshold for SKA-Low is found to be $ < 0.48"$, as displacements $> 0.048"$ cause significant deviations between the inferred astrophysical parameters and their actual values. This suggests that sky model position errors can bias the inference of the astrophysical parameters. 
     
    \item We also observed that, among all three parameters, $\zeta$ is the most sensitive to the power spectrum (PS). Even small variations in the PS, caused by calibration gain or position errors, have a significant impact on the ionization efficiency parameter, $\zeta$. 
    
\end{itemize}

 Thus, we conclude that for the upcoming SKA Low interferometer, the gain calibration error should be nearly 0.001\%, and the sky model position error should be less than 0.48" to avoid biasing the inference. Beyond these thresholds, these errors start to affect the results. To reduce their impact, calibration must be performed with the same level of accuracy, or we need to develop modeling or mitigation techniques to eliminate the residual foregrounds arising from calibration or sky model inaccuracies, ensuring minimal influence on the inference. It is important to note that the simulations presented in this study rely on several simplifying assumptions regarding foregrounds and instrumental effects. This underscores the need for more realistic and detailed simulations, including factors such as primary beam chromaticity and other instrumental systematics, to more accurately assess the robustness of parameter inference and to inform optimal array design. These complexities will be addressed in future work for a more comprehensive understanding.

\acknowledgments
AT would like to thank the Indian Institute of Technology Indore for providing funding for this study in the form of a Teaching Assistantship. SM and AD acknowledge financial support through the project titled ``Observing the Cosmic Dawn in Multicolour using Next Generation Telescopes'' funded by the Science and Engineering Research Board (SERB), Department of Science and Technology, Government of India through the Core Research Grant No. CRG/2021/004025. The authors acknowledge the use of facilities procured through the funding via the Department of Science and Technology, Government of India sponsored DST-FIST grant no. SR/FST/PSII/2021/162 (C) awarded to the DAASE, IIT Indore. The authors would like to thank the anonymous reviewer and the scientific editor for their thoughtful comments and helpful suggestions, which have contributed to improving the quality of this work. 
\appendix
\section{Robustness Evaluation of the Emulator-based MCMC Pipeline}\label{Robustness_check}

\begin{figure*}[h]
    \centering
    {\includegraphics[width=0.48\textwidth]{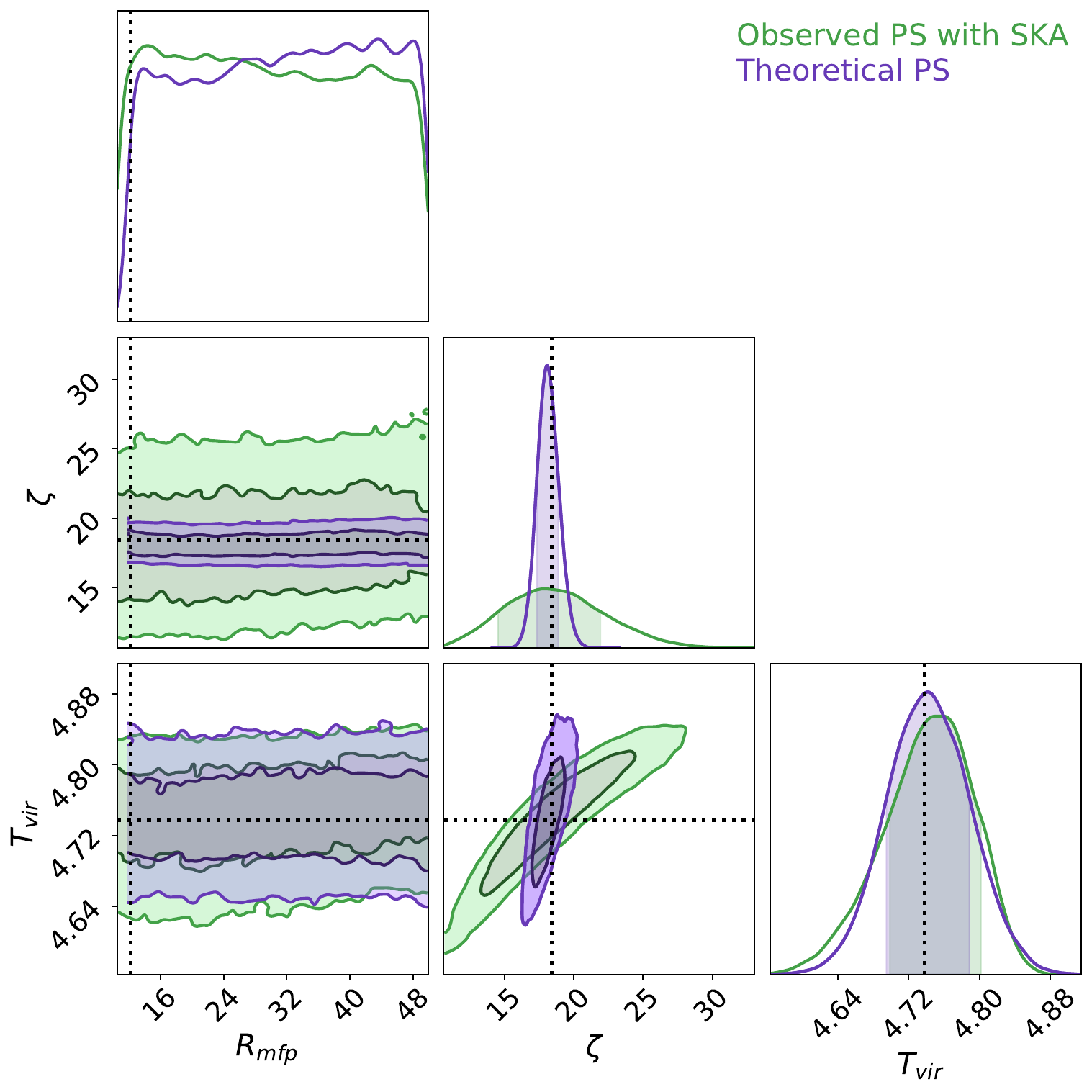}}
    {\includegraphics[width=0.48\textwidth]{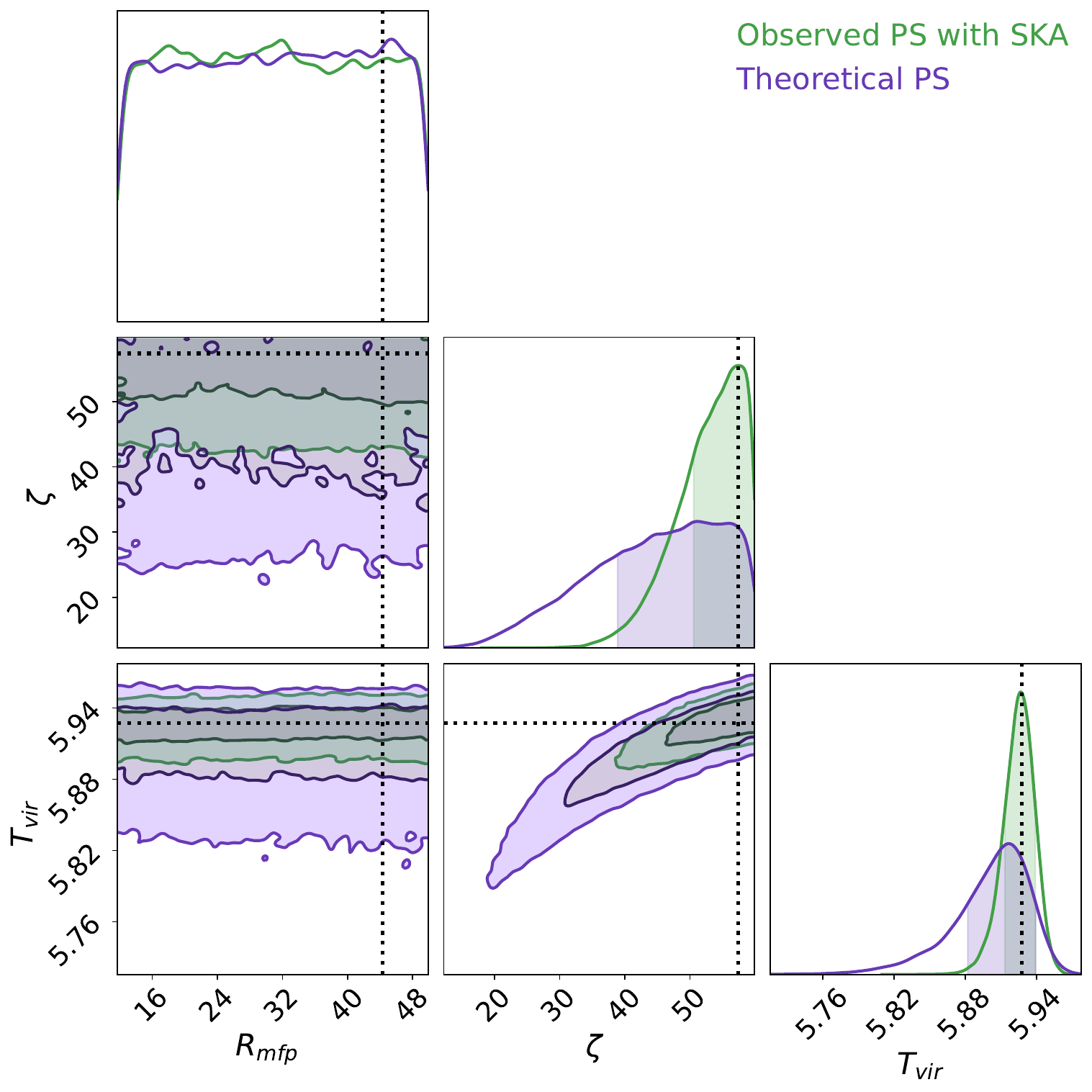}}
    \caption{Depicts the posterior distribution of model parameters obtained through power spectrum analysis, comparing the theoretical power spectrum with the observed power spectrum from SKA-Low for two different test sets (\textbf{left:} Set 2, \textbf{right:} Set 3). The enclosed areas between the inner and outer contours signify the 1$\sigma$ and 2$\sigma$ confidence levels, respectively.}
   
    \label{Fig_ps}
\end{figure*}

\begin{table}[!h]
\centering
\begin{tabular}{|c|c|c|c|}
\hline
\textbf{Set} & \begin{tabular}[c]{@{}l@{}}Parameters \\ \&\\ True value \end{tabular} & Theoretical PS & \begin{tabular}[c]{@{}l@{}}Observed PS \\ with SKA-Low\end{tabular} \\ \hline
 & $R_{mfp}$: 34.67 & \textbf{\textendash} & \textbf{\textendash} \\[5pt]
\textbf{Set 1}               & $\zeta$: 27.92  & $27.5^{+3.7}_{-3.4}$ & $27.76^{+7.5}_{-6.9}$  \\[5pt]
                   & $T_{vir}$: 5.428 & $5.428^{+0.023}_{-0.032}$ & $5.437^{+0.033}_{-0.046}$ \\[5pt] \hline
 & $R_{mfp}$: 12.2  & $50.0^{+0}_{-27}$ & \textbf{\textendash} \\[5pt]
\textbf{Set 2}               & $\zeta$: 18.42  & $18.09^{+0.80}_{-0.75}$ & $17.7^{+4.2}_{-3.2}$  \\[5pt]
               & $T_{vir}$: 4.738 & $4.734^{+0.046}_{-0.048}$ & $4.761^{+0.041}_{-0.062}$  \\[5pt] \hline
 & $R_{mfp}$: 44.33 & \textbf{\textendash}  & \textbf{\textendash}  \\[5pt]
\textbf{Set 3}               & $\zeta$: 57.42  & $50.8^{+9.0}_{-11.9}$ & $59.8^{+0.0}_{-9.3}$ \\[5pt]
               & $T_{vir}$: 5.928 & $5.917^{+0.022}_{-0.035}$ & $5.926^{+0.013}_{-0.013}$  \\[5pt] \hline
\end{tabular}
\caption{Inferred EoR model parameters ($\zeta$, $T_{vir}$, and $R_{mfp}$) with $1\sigma$ uncertainties for different parameter sets. Theoretical power spectrum (PS) estimates are compared with observed PS constraints from SKA-Low.}
\label{tab:inferred}
\end{table}

\bibliographystyle{JHEP}
\bibliography{biblio}
\end{document}